\documentclass[conference]{IEEEtran}
\ifCLASSINFOpdf
\else
\fi
\usepackage[cmex10]{amsmath}
\usepackage{amsmath}
\usepackage{amsfonts}
\usepackage{amssymb}
\usepackage{graphicx}
\usepackage{adjustbox}
\usepackage[T1]{fontenc}
\usepackage{caption}
\usepackage{subcaption}
\newcommand{\RNum}[1]{\uppercase\expandafter{\romannumeral #1\relax}}
\usepackage{mathtools}
\usepackage{array}
\newcolumntype{P}[1]{>{\centering\arraybackslash}p{#1}}
\begin{document}
%
\title{\textit{Removal of Narrowband Interference (PLI in ECG Signal) Using Ramanujan Periodic Transform (RPT)}}
\author{\IEEEauthorblockN{Basheeruddin Shah Shaik${^1}$, Vijay Kumar Chakka${^2}$, Srikanth Goli${^3}$, A. Satyanarayana Reddy${^4}$}
\IEEEauthorblockA{${^1{^,}^2{^,}^3}$Department of Electrical Engineering\\
${^4}$Department of Mathematics\\
Shiv Nadar University\\
Greater Noida, India\\
${^1}$bs600@snu.edu.in,${^2}$vijay.chakka@snu.edu.in,${^3}$gs499@snu.edu.in,${^4}$satyanarayana.reddy@snu.edu.in}
}
%


\maketitle

\begin{abstract}
Suppression of interference from narrowband frequency signals play vital role in many signal processing and communication applications. A transform based method for suppression of narrow band interference in a biomedical signal is proposed. As a specific example Electrocardiogram (ECG) is considered for the analysis. ECG is one of the widely used biomedical signal. ECG signal is often contaminated with baseline wander noise, powerline interference (PLI) and artifacts (bioelectric signals), which complicates the processing of raw ECG signal. This work proposes an approach using Ramanujan periodic transform for reducing PLI and is tested on a subject data from MIT-BIH Arrhythmia database. A sum ($E$) of Euclidean error per block ($e_i$) is used as measure to quantify the suppression capability of RPT and notch filter based methods. The transformation is performed for different lengths ($N$), namely $36$, $72$, $108$, $144$, $180$. Every doubling of $N$-points results in  $50{\%}$ reduction in error ($E$). 
\end{abstract}
\begin{IEEEkeywords}
 Ramanujan sum; RPT; PLI; Ramanujan Space; ECG; Notch filter.
\end{IEEEkeywords}
%
\IEEEpeerreviewmaketitle
\section{Introduction}
Interference caused in any system/signal by narrow band of frequencies known as narrowband interference. In the past several years there has been steady increase in research involving reduction of narrowband interference \cite{1094725}. The basic idea for any transformation technique is representing an input signal as a linear combination of orthogonal/linearly independent basis functions. A finite set of orthogonal complex exponential basis functions are used in finite length Fourier transform. A set of Ramanujan sums are used as basis functions for Ramanujan FIR Transform (RFT). Ramanujan sums and its linearly independent circular shifts forms basis functions for Ramanujan Periodic Transform (RPT) \cite{6839030}. One of the major advantage of RPT and RFT is having an integer basis.

Electrocardiogram (ECG) is an electrical record of depolarization and re-polarization activity of heart. Nobel laureate Willem Einthoven recorded an ECG signal in 1903 by placing electrodes on limbs and chest. Preprocessing of bio-potential signals such as ECG is necessary in applications like diagnosis of disease \cite{hosseini2006}, and QRS complex delineation for feature extraction\cite{1275572}, \cite{7383914}, to eliminate the contaminated artifacts, noises and interference. Powerline interference is an interference caused in ECG signal due to Electromagnetic interference (EMI) of frequency 50/60Hz from supply lines. The classical method for elimination of power line interference is passing the signal through a filter which reduces the interference, in the state of art, many methods exist to reduce the power line interference, few of them are based on adaptive filtering \cite{1451965}, using wavelets \cite{5670602}, based on state space recursive least squares filtering \cite{6530021}, and a well known method, design of adaptive and non-adaptive notch filter \cite{477707}. In this paper a method based on RPT is used to remove the power line interference.

In Section \RNum{2} a detailed description about Ramanujan sum, Ramanujan space and Ramanujan periodic transform is described. An overview of proposed methodology and simulation results are presented in section \RNum{3}. Summary of the overall paper and future work is discussed in Section \RNum{4}.

\section{RAMANUJAN PERIODIC TRANSFORM}
For a fixed positive integer $m$, the great Indian Mathematician Srinivasa Ramanujan introduced an trigonometrical summation, now called 
as Ramanujan sum \cite{Ramanujan}, denoted $S_m(n)$ and is defined as
\begin{equation}
\label{eq1}
 s_m(n) = \sum_{k=1,(k,m)=1}^{m}e^{\frac{j2{\pi}kn}{m}},
\end{equation}
where $(k,m)$ denotes the greatest common divisor (gcd) between $k$ and $m$. It is easy to check that $S_m(n)$ is a periodic and integer valued  function with period $m,$ thus it is a finite sequence.  For example, 
$S_3(n)=2,-1,-1.$
This attracts many signal processing researchers to use $s_m (n)$ as basis function. 

Let $\mathbb{N}$ and $\mathbb{C}$ denote the set of natural numbers (starting from $0$) and complex numbers respectively. Then a mapping 
$f:\mathbb{N}\to \mathbb{C}$ is called an arithmetic function. For example, for 
$n \in \mathbb{N}$, let $\phi(n)$ denote the number of positive integers not exceeding $n$
that are relatively prime to $n$. That is 
$\phi(n)=\#\{m|1\leq m\leq n, (m,n)=1\}.$  The arithmetic function $\phi(n)$ is called Euler's Phi-function or Euler's totient function.
Ramanujan's motivation for introducing this summation is to represent any arithmetic function as a linear combination of Ramanujan sums. In this paper we treat an arithmetic function as a sequence and if the arithmetic function is periodic, then it is a finite sequence.
For a given $m$ it is easy to verify $s_m(0) = \phi(m)$. Consider any two Ramanujan sums $s_{m_1}(n),s_{m_2}(n),$ they are orthogonal to each other over the sequence length $l$, where $l = lcm(m_1,m_2)$, i.e.,
\begin{equation}
\label{eq3}
\sum_{n=0}^{l-1}s_{m_1}(n)s_{m_2}(n-k) = 0,{\quad} {m_1}\neq{m_2},
\end{equation} for any integer $k$, where $0\leq k \leq {l-1}$.
 
\subsection{Ramanujan Space}
The column space of integer circulant matrix $D_m$ of size $m{\times}m$ forms the Ramanujan space $V_m$ \cite{6839014}. The first column of the matrix $D_m$ is the sequence $ s_m(n)$, the remaining $m-1$ columns are circular down shift of the previous columns. Let $m=3$, then,
\begin{equation}
D_3 = \begin{bmatrix}
s_3(0) & s_3(2) & s_3(1) \\
s_3(1) & s_3(0) & s_3(2) \\
s_3(2) & s_3(1) & s_3(0)
\end{bmatrix} = \begin{bmatrix}
2 & -1 & -1 \\
-1 & 2 & -1 \\
-1 & -1 & 2
\end{bmatrix}.
\end{equation}
From theorem 5 in \cite{6839014}, it is clear that any consecutive $\phi(m)$ columns of $D_m$ matrix forms the basis for $V_m$, but 
 not  an orthogonal basis unless $m$ is power of $2.$
Then any sequence $x(n)\in{V_m}$ can be expressed as follows,
\begin{equation}
x(n) = \sum_{k=0}^{\phi{(m)}-1}{\beta}_{k}s_m(n-k).
\end{equation}
It is shown in Theorem $10$ in \cite{6839014} that if we add two periodic signals  from Ramanujan space $V_m$, then the period of 
resultant signal is the LCM of periods of individual signals.  It is not true 
for any periodic signals, there the period of 
resultant signal is a divisor of  LCM of periods of individual signals.

If signals from different Ramanujan spaces are added i.e., 
\begin{equation}
x(n) = \sum_{{i}=1}^{k}{x_{m_i}(n)} ,{\quad}x_{m_i}(n)\in V_{m_i},
\end{equation}
then the period of $x(n)$ is equal to $N$, assuming that none of the $x_m (n)$ is identically zero (Theorem 12 in \cite{6839014}), where $N=lcm(m_1,m_2,\dotsc ,m_k)$. These are the two important properties of sequences belongs to Ramanujan spaces, are helpful for understanding RPT, as explained in subsection D.
\subsection{Relation between Ramanujan Spaces and DFT Matrix}
According to factorization property of $D_m$  \cite{6839014},
\begin{equation}
\label{eq:DM}
D_m=W W^H,
\end{equation}
where $W$ is an $m\times \phi(m)$ matrix obtained from the DFT matrix $F$ by choosing those columns such that the column numbers are relatively prime to $m.$ So the columns of $W$ acts as a basis for $V_m.$
If $x(n){\in}V_m$ then $x(n)$ can be expressed in any one of the following forms,
\begin{equation}
x(n) = \sum_{k=0}^{\phi{(m)}-1}{\beta}_{k}s_m(n-k) = \sum_{1{\leq}k{\leq}m,(k,m)=1} {\alpha_k}{e^\frac{-j2{\pi}kn}{m}},
\end{equation}
each term (in complex exponential summation) has period $m$, and the frequencies are,
\begin{equation}
\label{eq10}
w_k = \frac{2{\pi}k}{m},{\quad}{1{\leq}k{\leq}m},\quad {(k,m)=1}.
\end{equation}
So, \textit{\bfseries{Ramanujan space having period $\bold{m}$ has different frequency components}}.
\subsection{RPT}
In this transformation, a representation of input signal $x(n)$ of length $N$ as a linear combination of signals belongs to Ramanujan space $V_{m_i}$ (basis) is used \cite{6839030}, i.e.,
\begin{equation}
x(n) = \sum_{{m_i}|N}\sum_{k=0}^{\phi{(m_i)}-k}{\beta_i}_{k}s_{m_i}(n-k),
\end{equation}
where $m_i |N$ means the summation is executed for those $m_i$ values, which are divisors of $N$. Let us consider $N = 4$ length input sequence, then the divisors $(m_i)$ are $1$,$2$ and $4$, i.e., the input signal is represented as a linear combination of signals from Ramanujan spaces $V_1$,$V_2$, and $V_4$. 
\begin{equation}
\begin{aligned}
x(n) & = x_1 (n)  + x_2 (n)  + x_4 (n),\\ & \quad {x_1}(n) {\in} {V_1},{\quad} { x_2(n)} {\in}{V_2} {\quad}{ \&}{\quad} { x_4(n)} {\in} {V_4}.\\
 & = \sum_{l=0}^{\phi{(1)}-1}{\beta_1}_{l}s_{1}(n-l)+\sum_{l=0}^{\phi{(2)}-1}{\beta_2}_{l}s_{2}(n-l)\\
& +\sum_{l=0}^{\phi{(4)}-1}{\beta_4}_{l}s_{4}(n-l),{\quad} 0{\leq}n{\leq}3\\
& = \beta_{10}{s_1}(n)+\beta_{20}{s_2}(n)+\beta_{40}{s_4}(n)+\beta_{41}{s_4}(n-1)\\
x & = \underbrace{\begin{bmatrix}
{s_1}(0) & {s_2}(0) & {s_4}(0) & {s_4}(3)\\
{s_1}(0) & {s_2}(1) & {s_4}(1) & {s_4}(0)\\
{s_1}(0) & {s_2}(0) & {s_4}(2) & {s_4}(1)\\
\underbrace{{s_1}(0)}_{R_1} & \underbrace{{s_2}(1)}_{R_2} & \underbrace{{s_4}(3)}_{R_4} & \underbrace{{s_4}(2)}_{R_4}
\end{bmatrix}}_{T_4}
\underbrace{\begin{bmatrix}
\beta_{10}\\
\beta_{20}\\
\beta_{40}\\
\beta_{41}
\end{bmatrix}}_{\beta}\\ & = {[{R_1}\quad {R_2}\quad {R_4}]}{\beta}\\ & x = {T_4}{\beta}.
\end{aligned}
\end{equation}
This synthesis representation of $x(n)$ is known as Ramanujan Periodic Representation (RPR). In general for a given $N$ length sequence, the RPR is,
\begin{equation}
\label{RPR}
x(n) = {T_N}\beta(K),
\end{equation}
where the values of $\beta$ are known as Ramanujan coefficients and $T_N$ is the transformation matrix. From Orthogonality property (Equation  \eqref{eq3}) the column space of $R_{m_i}$ is orthogonal to $R_{m_k}$ for $i{\neq}k$, i.e.,
\begin{equation}
\label{orthogonal}
{R_{m_i}^H}{R_{m_k}} = 0,{\quad} k{\neq}i.
\end{equation}
The projection of input signal onto a specific space can be calculated using equation \eqref{Projection}.
\begin{equation}
\label{Projection}
x_{m_{i}} = \underbrace{R_{m_{i}}({R_{m_{i}}}^{H}R_{m_{i}})^{-1}{R_{m_{i}}}^{H}}_{{Projection}{\quad}{matrix}}x.
\end{equation}

Let $m_1$,$m_2$,$...$,$m_k$,$N$ are the possible divisors of $N$, then the sum of totient functions of all these divisors is equal to length of the signal \cite{6839030}, i.e.,
\begin{equation}
\Rightarrow \sum_{{m_i}|N}\phi(m_i) = N,
\end{equation}
from \eqref{RPR} the analysis equation can be written as
\begin{equation}
\label{RPT}
\beta(K) = {T_N}^{-1}x(n),
\end{equation}
this transformation is known as \textit{Ramanujan periodic transform (RPT)}, also equations \eqref{RPR} and \eqref{RPT} forms Ramanujan transformation pair. If $N$ is a power of $2$, then equation  \eqref{RPT} modified as,
\begin{equation}
\beta(K) = {T_N}^{T}x(n).
\end{equation}

\section{Proposed Method for Narrowband Interference Removal and Simulation Results}
\begin{figure}[!h]
\centering
 \includegraphics[width=3.5in,height=1.5in]{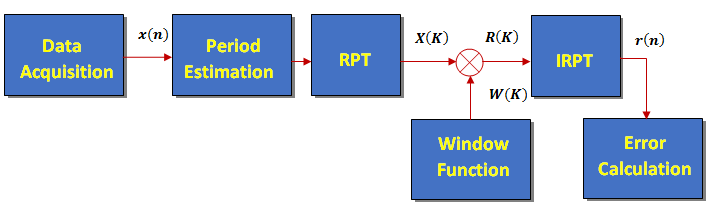} 
\caption[Block diagram of PLI reduction algorithm using RPT]{Block diagram of PLI reduction algorithm using RPT}
\label{f1}
\end{figure}

In the proposed methodology, a $N$ length narrowband interfered data ($x(n)$) is projected onto Ramanujan spaces, which are divisors of $N$. The window length is calculated based on the period of narrowband interference. 
A significant amount of projection energy in a Ramanujan space indicates the presence of periodic component of that space in the signal. In order to reduce the interference, first determine all periods of narrowband interference, then identify the matched Ramanujan spaces and the coefficients ($\beta(K)$) belongs to these spaces. Then force these coefficients to zero, this can be achieved by multiplying with a window function ($W(K)$), whose values are zeros for these spaces otherwise one. Then reconstruct (using equation \eqref{RPT}) the signal ($r(n)$) using these modified Ramanujan coefficients (${\beta(K)}{W(K))}$. As an example, the entire procedure described above is applied on ECG signal to remove the $50/60$Hz powerline interference. Block diagram of the proposed methodology for analysing ECG signal is shown in Fig. \ref{f1}. 

The brief description of each stage is presented  in the following sub-sections.
\subsection{Data Acquisition}
\begin{figure}[!h]
\centering
\includegraphics[width=3.5in,height=1.6in]{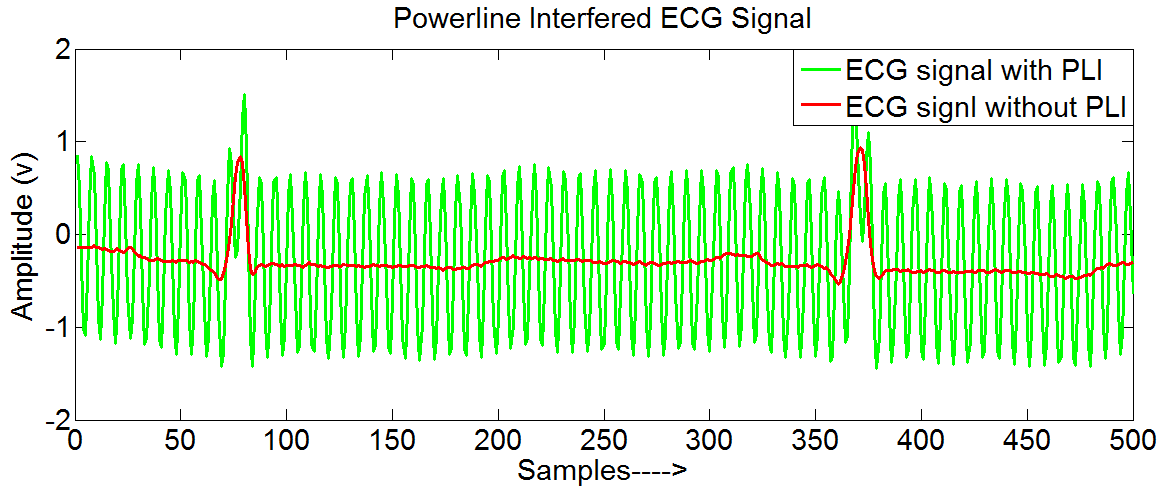} 
\caption[Powerline interfered ECG signal]{Powerline interfered ECG signal}
	\label{Powerline interfered ECG signal}
\end{figure}

MIT-BIH Arrhythmia database from physionet website \cite{Physionet} is used for the analysis it consists of two channel ECG recordings of total $48$ subjects, with a sampling frequency of $360$Hz. 
The proposed methodology is analysed on a single record. To obtain a PLI ECG data a pure sinusoidal signal of $50$Hz frequency is added to the entire record. A typical powerline interfered ECG signal (Record 100 in MIT-BIH Arrhythmia) up to 500 samples is shown in Fig.\ref{Powerline interfered ECG signal}.   
\subsection{Period Estimation}
To estimate the period of an unknown signal, one need to decompose the signal into different Ramanujan spaces. Then it is easy to estimate the period by calculating projection energies ($|x_{m_i}|^2$). Since PLI is a $50$Hz sinusoid, with the sampling frequency of $360$Hz, corresponds to a period of $36$ (with minimum possible integer value of $k = 5$ in  \eqref{eq10}), 
\begin{figure}[!h]
\centering
 \includegraphics[width=3.5in,height=2.2in]{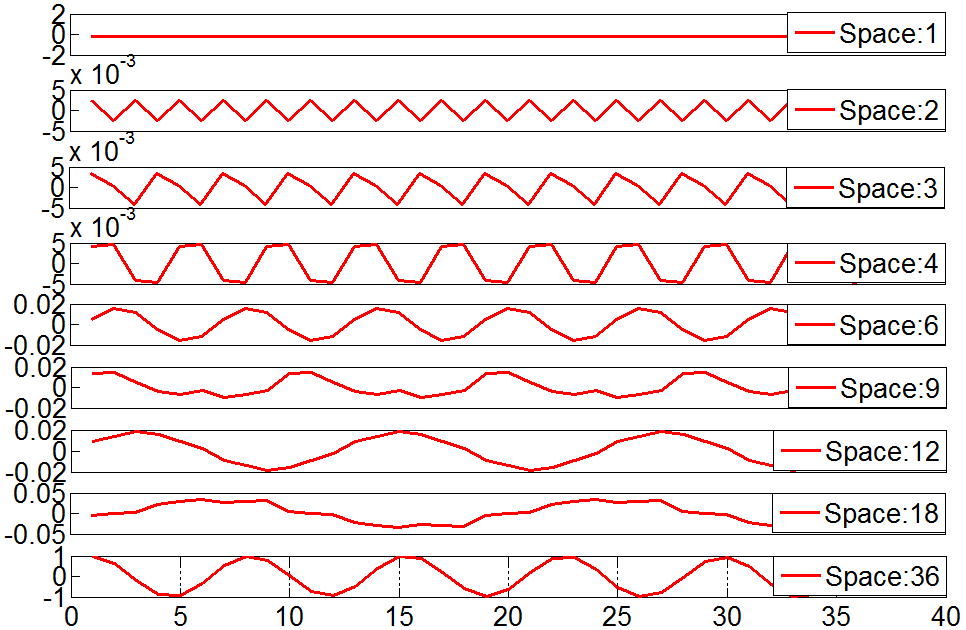} 
\caption[Decomposition of ECG signal with PLI into different periodic signals corresponding to their respective Ramanujan spaces]{Decomposition of ECG signal with PLI into different periodic signals corresponding to their respective Ramanujan spaces}
\label{Decomposition of ECG signal with PLI into different periodic signals corresponding to their respective Ramanujan spaces}
\end{figure}

\subsection{Implementation of RPT}
Now the entire signal should be divided in to number blocks such that period $36$ should be one of the divisors of block size. So, the minimal possible block size is $36$. In the current analysis a block size of $36$ is considered. The projections $x_{m_i}$ on each $m_i$ (which is a divisor of block size) is calculated using \eqref{Projection}. Figure \ref{Decomposition of ECG signal with PLI into different periodic signals corresponding to their respective Ramanujan spaces} shows the decomposition of ECG signal with PLI into different periodic signals corresponding to their respective Ramanujan spaces.

Space wise normalized RPT is implemented on each block of data. To calculate normalized RPT ($\hat{T_N}$), the following procedure is considered. Let $N = 4$, then, 
\begin{equation}
\label{matrix}
T_4 = \begin{bmatrix}
1 & 1 & 2 & 0\\
1 & -1 & 0 & 2 \\
1 & 1 & -2 & 0\\
1 & -1 & 0 & -2
\end{bmatrix}.
\end{equation}
According to \eqref{matrix}, the multiplication yields to
\begin{equation}
{T_4}^{T}{T_4} = \begin{bmatrix}
4 & 0 & 0 & 0\\
0 & 4 & 0 & 0 \\
0 & 0 & 8 & 0\\
\underbrace{0}_{V_1} & \underbrace{0}_{V_2} & \underbrace{0}_{V_4} & \underbrace{8}_{V_4}
\end{bmatrix}.
\end{equation}
The diagonal elements in the above matrix representing the space wise energy of $N$ length Ramanujan sum with period $m$ (where $m$ is a divisor of $N$). To overcome this issue a space wise normalization is performed, for example the first column ($\in{V_1}$) in $T_4$ is divided with $\sqrt{4}$, second column ($\in{V_2}$) is divided with $\sqrt{4}$, third and fourth ($\in{V_4}$) are divided with $\sqrt{8}$. By using above normalized $\hat{T_4}$, the multiplication yields to, 
\begin{equation}
\small{{\hat{T_4}}^{T}{\hat{T_4}}=
\begin{bmatrix}
1 & 0 & 0 & 0\\
0 & 1 & 0 & 0 \\
0 & 0 & 1 & 0\\
0 & 0 & 0 & 1
\end{bmatrix}},
\end{equation} 
input signal $x(n)$ is transformed by this $\hat{T_N}$ matrix to generate the Ramanujan periodic coefficients ($\beta(K)$). These are manipulated to reduce the PLI.

\subsection{Frequency Domain Analysis and Implementation of IRPT}
\begin{figure}[!h]
\centering
 \includegraphics[width=3.5in,height=1.6in]{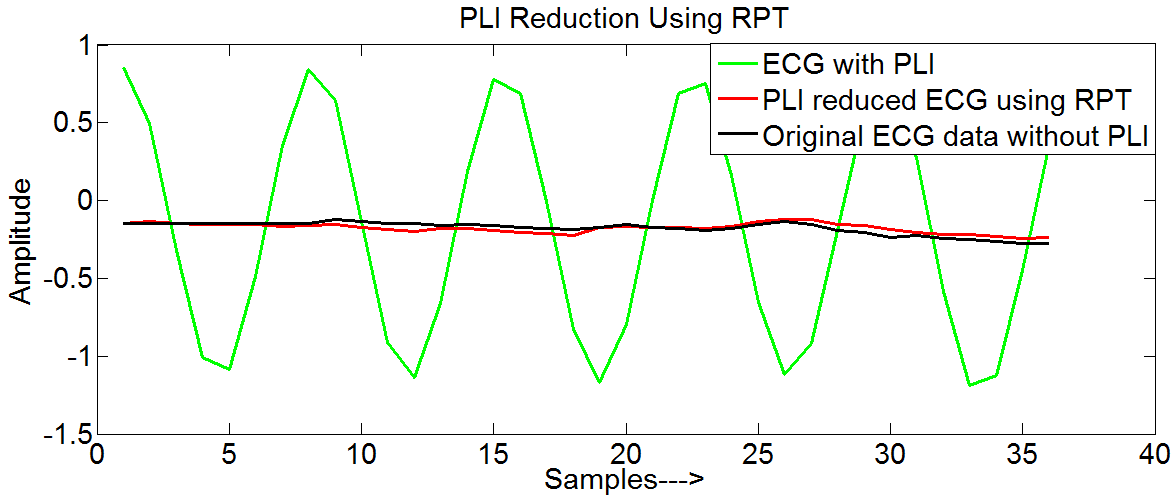} 
\caption[PLI reduction using RPT]{PLI reduction using RPT}
\label{PLI reduction using RPT}
\end{figure}

Now the obtained $36$ Ramanujan coefficients ($\beta$) should be separated space wise.
The last $12$ coefficients in $36$ belongs to $V_{36}$. In order to reduce PLI, the coefficients in $V_{36}$ should be equal to zero.  For this purpose a window sequence $(W(K))$ of length $36$ is defined as,
\begin{equation}
\label{window}
W(K) = \{\underbrace{1}_{0},\underbrace{1}_{1},\underbrace{1}_{2},\cdots,\underbrace{1}_{23},\underbrace{0}_{24},\cdots,\underbrace{0}_{35}\},
\end{equation}
which is multiplied with  $\beta(K)$, resulting in modified Ramanujan coefficients (${\beta(K)}W(K)$) are reconstructed back as signal ($r(n)$) by using  Ramanujan periodic representation  \eqref{RPR} known as Inverse Ramanujan periodic transform (IRPT). The reconstructed (PLI reduced) ECG data for a particular block of $36$ samples is shown in 
\ref{PLI reduction using RPT}. The magnitude response of ECG signal with PLI, before and after applying RPT is shown in Fig. \ref{Magnitude Response of original ECG Signal and With, Without (reduced using RPT) PLI} 

\begin{figure}[!h]
\centering
 \includegraphics[width=3.5in,height=1.6in]{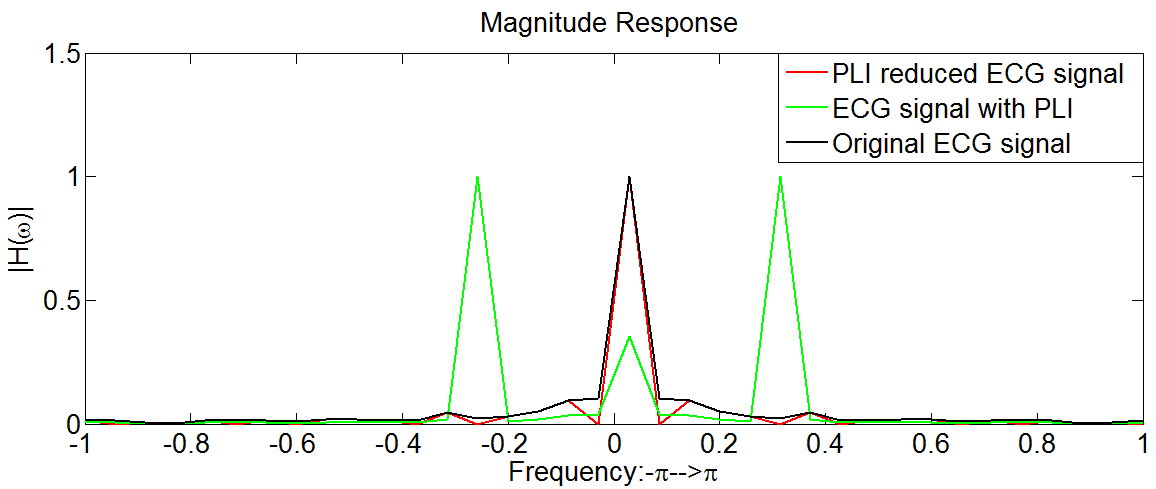} 
\caption[Magnitude Response of original ECG Signal and With, Without (reduced using RPT) PLI]{Magnitude Response of original ECG Signal and With, Without (reduced using RPT) PLI}
\label{Magnitude Response of original ECG Signal and With, Without (reduced using RPT) PLI}
\end{figure}

\subsection{PLI Reduction using Notch Filter}
\begin{figure}[!h]
\centering
 \includegraphics[width=3.5in,height=1.6in]{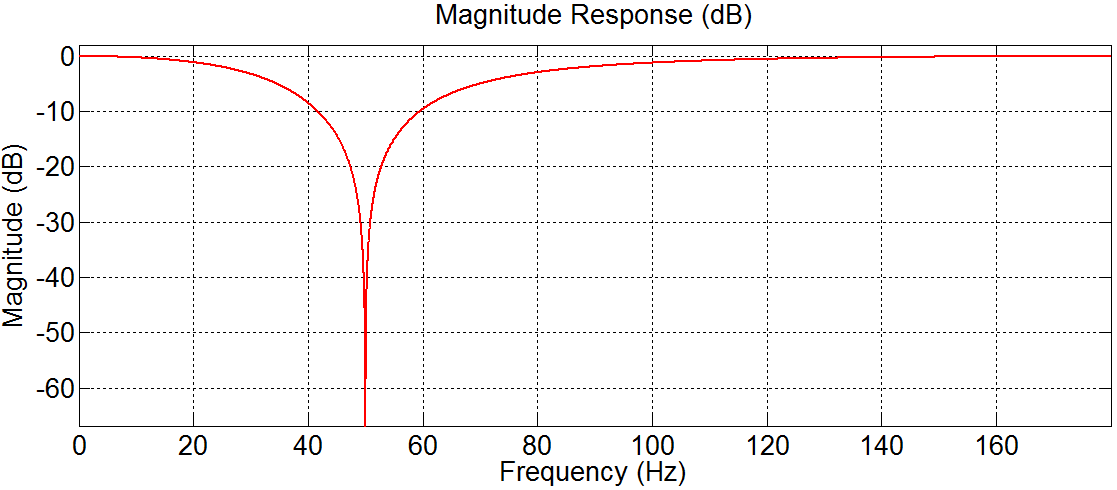} 
\caption[Magnitude Response Notch Filter]{Magnitude response notch filter}
\label{Magnitude Response Notch Filter}
\end{figure}

The results of proposed algorithm are compared with the results obtained using notch filter, by using Euclidean error as a measure. The relation between notch frequency ($f_0$), quality factor ($Q$), and bandwidth ($\delta{f}$) is mathematically represented in equation \eqref{notch} \cite{6530021}.
\begin{equation}
\label{notch}
Q = \frac{f_0}{\delta{f}}.
\end{equation}
Notch filter with a higher attenuation level will remove the PLI noise very effectively, to achieve higher attenuation the $Q$ factor should be decreased. In this work a second order IIR notch filter with notch frequency of $50$Hz, having a quality factor of $1$ ($\delta{f} = 50$Hz) is designed. Fig. \ref{Magnitude Response Notch Filter}. shows the magnitude response of notch filer.

\begin{figure}[!h]
\centering
 \includegraphics[width=3.5in,height=1.6in]{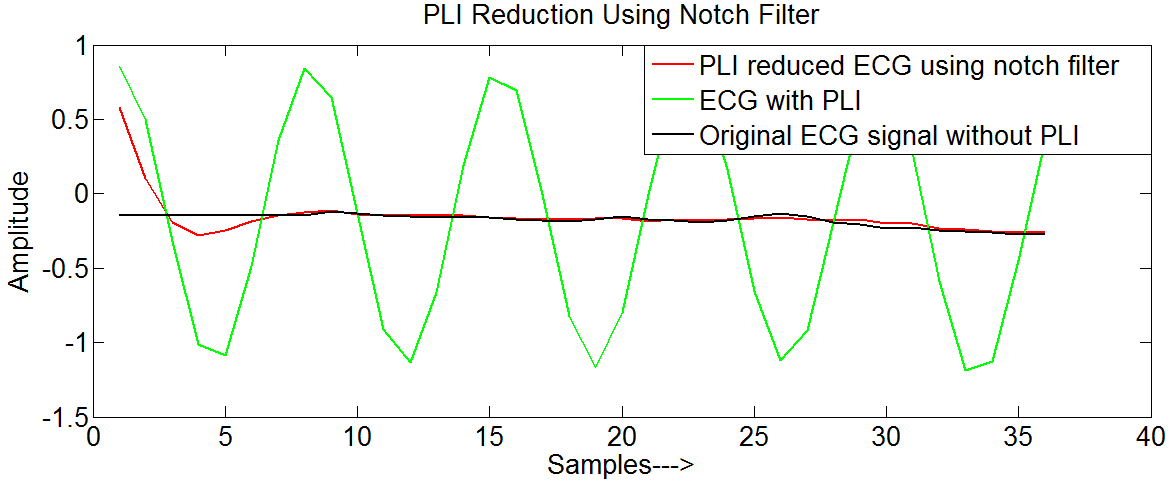} 
\caption[PLI Reduction Using Notch Filter]{PLI reduction using notch filter}
\label{PLI Reduction Using Notch Filter}
\end{figure}

\begin{figure}[!h]
\centering
 \includegraphics[width=3.5in,height=1.6in]{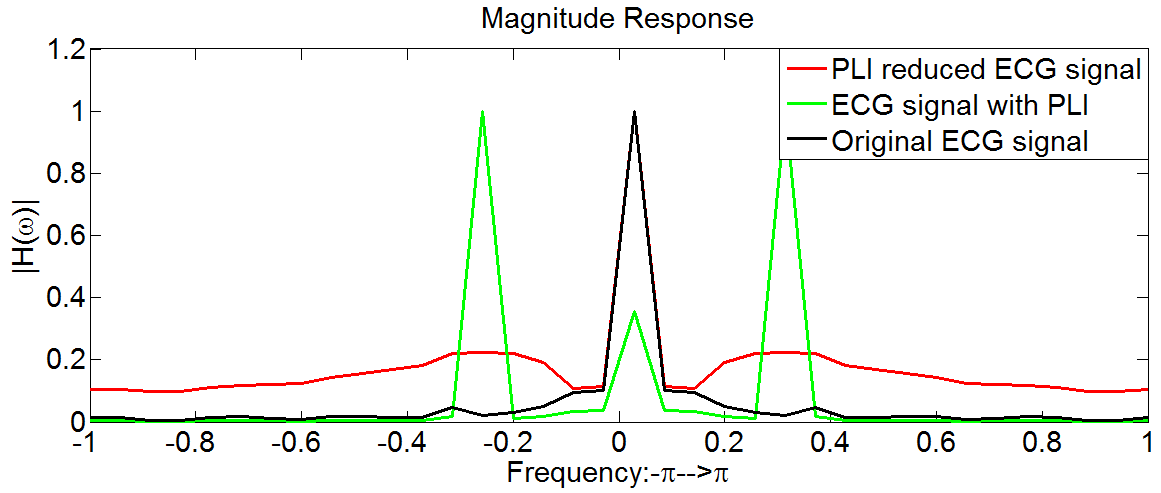} 
\caption[Magnitude Response PLI Reduced ECG Signal Using Notch Filter]{Magnitude response PLI reduced ECG signal using notch filter}
\label{Magnitude Response PLI Reduced ECG Signal Using Notch Filter}
\end{figure}

The analysis using notch filter is also performed block by block (similar to RPT), so the reduction of PLI in the first block of data using notch filter is shown in Fig. \ref{PLI Reduction Using Notch Filter}.
The magnitude response of the PLI reduced ECG signal using notch filter is shown in Fig.\ref{Magnitude Response PLI Reduced ECG Signal Using Notch Filter}.

\subsection{Euclidean Error Measurement}
For each block of data an error is calculated by using the equation \eqref{Error}.
\begin{equation}
\label{Error}
Error (e_i) = \sum_{n=0}^{N-1}|{x_i}(n)-{r_i}(n)|^2.
\end{equation}
Where ${x_i}(n)$ is the $i^{th}$ block input ECG signal with out any interference, ${r_i}(n)$ is the $i^{th}$ block interference reduced signal and $N$ is the block size. For the first block of data ${e_i}$ using RPT method is 0.0268, using Notch filter is 0.6282.

A length of $524288$ samples of ECG data is considered from the record 100 of MIT-BIH database, and it is converted to $36\times14564$ ($14564\times36 = 524304$, by appending few zeros to the considered signal) after deciding the block size. Now ${e_i}$ obtained using two methods is calculated for a total of $14564$ blocks. Then an $E$, defined in \eqref{E}, is calculated for the total blocks.
\begin{equation}
\label{E}
E = \sum_{i=1}^{14564}{e_i}.
\end{equation} 
The above procedure is repeated for different window lengths (different $N$-point RPT's).For each length, an $E$ is calculated, and summarized in TABLE
\ref{tab:Comparison of total error between RPT and Notch methods for different block sizes}.
\begin{table}[h]
\centering
\caption{Comparison of $E$ between RPT and Notch methods for different block sizes}
\label{tab:Comparison of total error between RPT and Notch methods for different block sizes}
\begin{adjustbox}{max width=\textwidth}
\renewcommand{\arraystretch}{1.25}
\begin{tabular}{|P{2.5cm}|P{2.5cm}|P{2.5cm}|}\hline 
\bfseries{\small{Block Size}}	&	 \bfseries{\small{RPT}}	&	\bfseries{\small{Notch Filtering}} \\    \hline
\small{36}	&	\small{7280}	&	\small{13652}	\\	\hline
\small{72}	&	\small{3501}	&	\small{8119}	\\	\hline
\small{108}	&	\small{2369}	&	\small{6266}	\\	\hline
\small{144}	&	\small{1782}	&	\small{5347}	\\	\hline
\small{180}	&	\small{1440}	&	\small{4805}	\\	\hline
\end{tabular}
\end{adjustbox}
\end{table}

\section{Conclusion}
In this work, a transform domain analysis based reduction of narrowband interference has been proposed using RPT. From TABLE \ref{tab:Comparison of total error between RPT and Notch methods for different block sizes}, it has been shown that RPT is reducing PLI with minimum $E$, in comparison with notch filter technique for different data window lengths. Every doubling of data length results in approximately $50\%$ reduction in error ($E$) for the case of RPT based method.
\section*{Acknowledgement}
The authors would like to thank Dr. Krishnan Rajkumar for his continuous support and many useful discussions.


%

\end{document}